\documentclass[preprint2]{aastex}

\newcommand{\oversim}[2]{\protect{\mbox{\lower0.5ex\vbox{%
  \baselineskip=0pt\lineskip=0.2ex
  \ialign{$\mathsurround=0pt #1\hfil##\hfil$\crcr#2\crcr\sim\crcr}}}}}
\newcommand{\simgreat}{\mbox{$\,\mathrel{\mathpalette\oversim>}\,$}} 
\newcommand{\simless} {\mbox{$\,\mathrel{\mathpalette\oversim<}\,$}} 

\slugcomment{ApJ, submitted 23.07.03, accepted 13.08.03}
\shorttitle{Field-IMFs of Massive Stars}
\shortauthors{Kroupa \& Weidner}

\begin{document}


\title{Galactic-Field IMFs of Massive Stars}


\author{Pavel Kroupa\altaffilmark{1} and Carsten Weidner}
\affil{Institut f\"ur Theoretische Physik und Astrophysik,
  Universit\"at Kiel, D-24098 Kiel, Germany}
\email{pavel/weidner@astrophysik.uni-kiel.de\\(submitted 27th July,
  accepted by ApJ 13th August 2003)}


\altaffiltext{1}{Heisenberg Fellow}

\begin{abstract}
Over the past years observations of young and populous star clusters
have shown that the stellar IMF appears to be an invariant featureless
Salpeter power-law with an exponent $\alpha=2.35$ for stars more
massive than a few~$M_\odot$. A consensus has also emerged that most,
if not all, stars form in stellar groups and star clusters, and that
the mass function of young star clusters in the solar-neighborhood and
in interacting galaxies can be described, over the mass range of a few
$10\,M_\odot$ to $10^7\,M_\odot$, as a power-law with an exponent
$\beta\approx2$. These two results imply that galactic-field IMFs for
early-type stars cannot, under any circumstances, be a Salpeter
power-law, but that they must have a steeper exponent $\alpha_{\rm
field}\simgreat 2.8$. This has important consequences for the
distribution of stellar remnants and for the chemo-dynamical and
photometric evolution of galaxies.
\end{abstract}


\section{Introduction}
\label{sec:intro}

For stars more massive than the Sun the stellar initial mass function
(IMF) can be approximated well by a single power-law function,
$\xi(m)\propto m^{-\alpha}$, with the Salpeter index $\alpha=2.35$,
$\xi(m)\,dm$ being the number of stars in the mass interval $m,m+dm$.
Massey and collaborators (Massey et al. 1995a, 1995b; Massey \& Hunter
1998; Parker et al. 2001; Massey 2002) have shown that $\alpha$ and
the mass of the most massive star, $m_{\rm max}$, in a cluster or OB
association are invariant in the metalicity range $0.002 \simless
Z\simless 0.02$, with $m_{\rm max}$ only depending statistically on
the richness of the cluster or association. The Salpeter-power law
form describes the distribution of stellar masses down to
approximately $0.5\,M_\odot$, where the IMF flattens to
$\alpha_1\approx 1.3$ (Kroupa et al. 1993; Reid et al. 2002), with a
further flattening near the sub-stellar mass limit (Chabrier 2003).

Stars form in clusters that contain from a dozen or so members to many
millions of stars. The mass-distribution of local embedded clusters is
a power-law, $\xi_{\rm ecl}\propto M_{\rm ecl}^{-\beta}$, with $\beta
\approx 2$ for $20 \simless M_{\rm ecl}/M_\odot \simless 1100$ (Lada
\& Lada 2003). Hunter et al. (2003) find $\beta=2-2.4$ for the initial
cluster mass function in the Large and Small Magellanic Clouds, and
Zhang \& Fall (1999) report $\beta=2\pm0.08$ for young star clusters
with $10^4 \simless M_{\rm ecl}/M_\odot \simless 10^6$ in the Antennae
galaxies. The consensus thus emerges that the cluster mass function
(CMF) of young star clusters of all masses can be described by a
single power-law form with $\beta\approx2$, which is consistent with
the star-cluster samples in six nearby spiral galaxies studied by
Larsen (2002) as well as the MF of young globular clusters (Richtler
2003).

The stellar population of a galaxy is progressively build-up over time
through star-formation in clusters. Low-mass clusters are much more
abundant than massive clusters and contribute most of the stars but do
not contain massive stars. Consequently the field-star IMF depends on
the form of the MF of star clusters, and will be steeper ($\alpha_{\rm
field} > \alpha$) than the stellar IMF in an individual massive star
cluster.  The aim of this contribution is to point out this important
consequence of clustered star formation on the field-star IMF.

Section~\ref{sec:meth} quantifies this effect for different cluster
and stellar MFs and Section~\ref{sec:concs} presents the discussion
and conclusions.

\section{The Field IMF from Clustered Star Formation}
\label{sec:meth}

The composite, galactic-field IMF is obtained by summing up the
stellar IMFs contributed by all the star clusters that formed over the
age of a galaxy,
\begin{equation}
\xi_{\rm field}(m) = \int_{M_{\rm ecl,min}}^{M_{\rm ecl,max}} \xi(m\le
                      m_{\rm max})\, \xi_{\rm ecl}(M_{\rm ecl})\,dM_{\rm ecl},
\label{eq:imfsum}
\end{equation}
where $\xi_{\rm ecl}$ is the MF of embedded clusters and $\xi(m\le
m_{\rm max})$ is the stellar IMF in a particular cluster within which
the maximal stellar mass is $m_{\rm max}$. The mass of the most massive
star in an embedded cluster with stellar mass $M_{\rm ecl}$ is given
by
\begin{equation}
1 = \int_{m_{\rm max}}^{m_{\rm max*}} \xi(m)\,dm,
\label{eq:mm}
\end{equation}
with
\begin{equation}
M_{\rm ecl} = \int_{m_{\rm l}}^{m_{\rm max}} m\,\xi(m)\,dm.
\label{eq:Mecl}
\end{equation}
On combining eqs.~\ref{eq:mm} and~\ref{eq:Mecl} the function $m_{\rm
max} = {\rm fn}(M_{\rm ecl})$ is quantified by Weidner \& Kroupa
(2003) who infer that there exists a fundamental upper stellar mass
limit, $m_{\rm max*}\approx150\,M_\odot$, above which stars do not
occur, unless $\alpha\simgreat3$ in which case no conclusions can be drawn
based on the expected number of massive stars. We thus have, for each
$M_{\rm ecl}$, the maximal stellar mass, $m_{\rm max}\le m_{\rm
max*}$, and with this information eq.~\ref{eq:imfsum} can be evaluated
to compute the field-star IMF.

\begin{figure}
\plotone{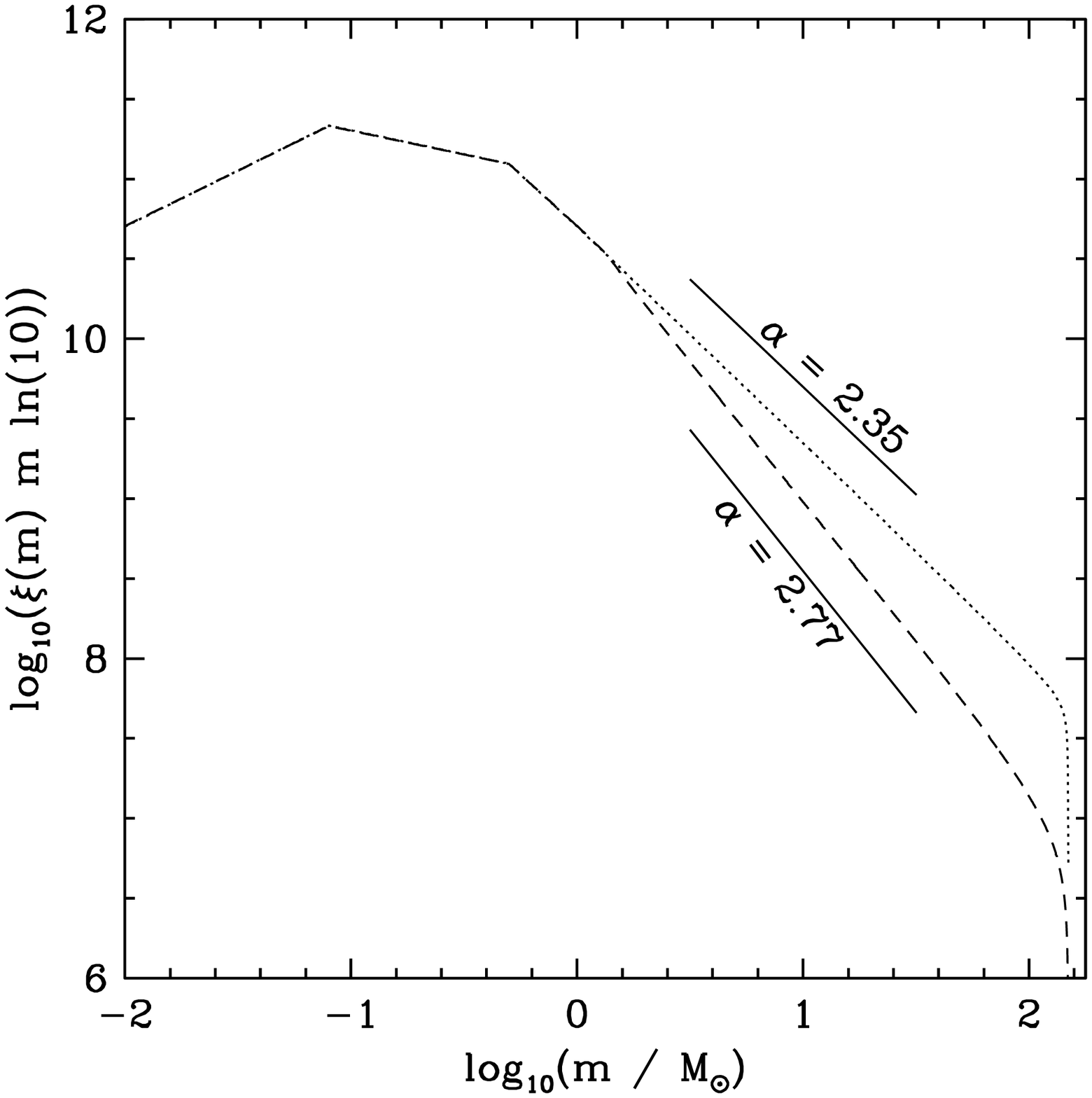}
\vspace{-20mm}
\caption{The dotted line is the standard stellar IMF, $\xi(m)$, in
  logarithmic units and given by the standard four-part power-law form
  (eq.~10 in Weidner \& Kroupa 2003), which has $\alpha=2.35$ for
  $m>1\,M_\odot$. The dashed line is $\xi_{\rm field}(m)$ for
  $\beta=2.2$.  The IMFs are scaled to have the same number of objects
  in the mass interval $0.01-1.0\,M_\odot$.  Note the turndown near
  $m_{\rm max*}=150\,M_\odot$ which comes from taking the fundamental
  upper mass limit explicitly into account (Weidner \& Kroupa 2003).
  Two lines with slopes $\alpha_{\rm line}=2.35$ and $\alpha_{\rm
  line}=2.77$ are indicated.
\label{fig:ex1}}
\end{figure}

Fig.~\ref{fig:ex1} shows the result assuming $\beta=2.2$ for the
clusters and $\alpha=2.35$ for the stars in each cluster. For the
lower ``stellar'' mass limit $m_{\rm l}=0.01\,M_\odot$ is adopted,
while the stellar IMF has, in all cases the standard (or
``universal'') four-part power-law form, $\alpha_0=+0.3
(0.01-0.08\,M_\odot)$, $\alpha_1=+1.3 (0.08-0.5\,M_\odot)$, and
$\alpha_2=+2.3 (0.5-1\,M_\odot)$ (Kroupa 2001).  For the minimum
``cluster'' mass, $M_{\rm ecl,min}=5\,M_\odot$ (a dozen stars), and
for the maximal cluster mass, $M_{\rm ecl,max}=10^7\,M_\odot$, are
used.  The power-law index, $\alpha_{\rm field}$, of the field-star
IMF is calculated from $\xi_{\rm field}$ at log$_{10}(m/M_\odot)=0.5$
and 1.5.  The resulting field-star IMF is well approximated by
$\alpha_{\rm field}=2.77$ for $m\simgreat 1\,M_\odot$.  Most massive
stars have more than one companion (Zinnecker 2003).  It can be shown
(Sagar \& Richtler 1991) that the binary-star corrected $\alpha$ is
larger by about~0.4 for a binary fraction of 100~per cent.  If the
binary-star-corrected IMF in each cluster has $\alpha=2.7$ then the
resulting field-star IMF is even steeper with $\alpha_{\rm
field}\approx3.2$.  The steepening of the field-star IMF as a function
of the star-cluster MF is plotted in Fig.~\ref{fig:avsb} for three
different values of $\alpha$. It becomes increasingly pronounced the
larger $\alpha$ is.  Below about $1\,M_\odot$, $\xi_{\rm field}$ and
$\xi$ have the same shape (i.e. the same $\alpha_0, \alpha_1,
\alpha_2$) because the minimum ``cluster'', $M_{\rm
ecl,min}=5\,M_\odot$, only contains stars with masses less than
$m_{\rm max}=1.3\,M_\odot$ (Weidner \& Kroupa 2003).

\section{Discussion and Conclusions}
\label{sec:concs}

Field-star IMFs must therefore always be steeper for $m>1\,M_\odot$
than the stellar IMF that results from a local star-formation event
such as in a star cluster.  This has rather important
implications. For example, the number and luminosity function of white
dwarfs and galactic supernova rates are determined by the field-star
IMF and are rarer than predicted by a stellar IMF
(Fig.~\ref{fig:rat}). For a realistic cluster MF with $\beta=2.2$ the
SN rate is 2.5--5 times smaller than for a stellar IMF.  The chemical
evolution and global energy feedback into the interstellar medium of a
galaxy is determined by the total number of stars born in a
distribution of star clusters, and thus by $\xi_{\rm field}$ rather
than $\xi$. Because $\xi_{\rm field}(m) < \xi(m)$ for $m> 1\,M_\odot$
a smaller effective yield and less feedback energy results per
low-mass star. Stellar mass-to-light ratia of galaxies are often
calculated by assuming the underlying IMF is a single Salpeter
power-law, $\xi_{\rm Salp}$, over the mass interval $0.1-100\,M_\odot$
(McCaugh et al. 2000).  For a given galaxy luminosity a Salpeter
power-law IMF overestimates the mass that has been assembled in stars
by a factor of about $4/3$.  Calculating $M_{\rm field} = X\,M_{\rm
Salp}$, where $M_{\rm field}$ is the total mass ever to have been in
field stars and $M_{\rm Salp}$ the corresponding mass for $\xi_{\rm
Salp}$, in the mass interval $0.1-100\,M_\odot$, and scaling $\xi_{\rm
field}$ and $\xi_{\rm Salp}$ to have the same number of stars in the
mass interval $1-100\,M_\odot$, it follows that $X(\alpha_{\rm
field})$ has an approximately parabolic shape with $X=0.73,
\alpha_{\rm field}=2.35$, a minimum value $X=0.72$ at $\alpha_{\rm
field}=2.6$ and $X=0.83, \alpha_{\rm field}=3.5$, in fine agreement
with the dynamical mass estimates of McCaugh (2003).  Note that
brown-dwarfs contribute less than 7\% by mass to a zero-age field-star
population for $\alpha_{\rm field} < 4.5$ (Kroupa 2002).  Furthermore,
$\alpha_{\rm field}(m>1\,M_\odot)$ must vary from galaxy to galaxy,
because the star-cluster MF varies in dependence of the star-formation
rate (SFR) in the sense of a correlation between $M_{\rm ecl,max}$ and
SFR (Larsen 2002). This is being quantified by Weidner, Kroupa \&
Larsen (in preparation), and implies that a complete description of
the chemo-dynamical and photometric evolution of galaxies needs the
three functions $\xi(m)$, $\xi_{\rm ecl}(M_{\rm ecl})$, and the
star-formation history.

In practice the field-star IMF cannot be observed directly for
$m\simgreat 1\,M_\odot$ because these stars have main-sequence
life-times shorter than the age of a galaxy. Salpeter (1955) estimated
$\xi_{\rm field}(m)$ from the luminosity function of nearby field
stars by correcting the star-counts for stellar evolution and assuming
a uniform stellar birth-rate and found $\alpha_{\rm Salp}=2.35$ for
$0.4<m/M_\odot<10$. More recently (and with improved stellar evolution
theory) Scalo (1986) and Reid et al. (2002) estimated a field-star IMF
for the Milky Way (MW) by first constructing the present-day MF by
counting massive stars in the local volume and applying corrections
for stellar evolution, the star-formation history of the MW and
diffusion of stellar orbits. The resulting field-star IMF can be
approximated for $m>1\,M_\odot$ with $\alpha_{\rm Scalo}=2.7$ (Kroupa
et al. 1993), $\alpha_{\rm Reid}=2.5-2.8$.  Yuan (1992) constrained
the field-star IMF from the distribution of stellar remnants, and
calculating $\alpha_{\rm field}$ from $\xi_{\rm Yuan}(m)$ at
log$_{10}(m/M_\odot)=0.2$ and 1.2, the Yuan-field-IMF has $\alpha_{\rm
Yuan}=2.8$ (Yuan's fig.15c) and $\alpha_{\rm Yuan}=2.7$ (Yuan's
fig.16c).  These estimates are consistent with $\alpha=2.35$ and
$\beta=2.2$, but they rely on assumptions concerning stellar evolution
and the stellar birth-rate history of the MW. Consequently the more
direct measurement of $\alpha$ from very young clusters (yielding
$\alpha_{\rm field}\approx2.3$) was preferred more recently to define
the ``standard'', or ``Galactic-field IMF'' (Kroupa 2001,
2002). However, this contribution has made it apparent that
$\alpha_{\rm field}\simgreat2.7$ is closer to the truth.  It therefore
appears that the field-star IMF should show a change in index at
$1<m_1/M_\odot<10$ from $\alpha=2.3$ ($m < m_1$) to
$\alpha=\alpha_{\rm field}$ ($m\ge m_1$).

\begin{figure}
\plotone{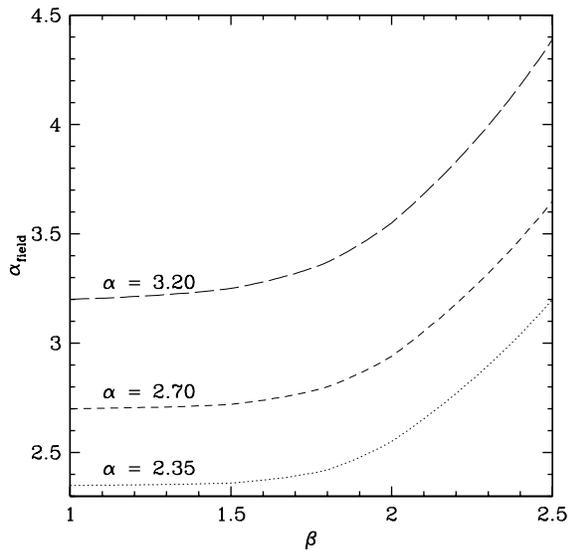}
\vspace{-25mm}
\caption{The field-star power-law index $\alpha_{\rm field}$
($m>1\,M_\odot$) as a function of the star-cluster MF power-law index
$\beta$ for $\alpha=2.35, 2.7, 3.2$. The slope $\alpha_{\rm field}$ is
calculated as in \S~\ref{sec:meth}.
\label{fig:avsb}}
\end{figure}

\begin{figure}
\plotone{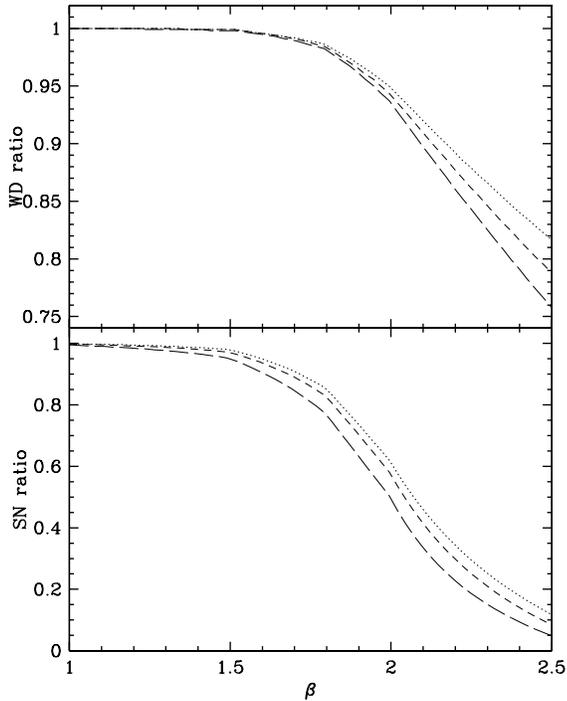}
\vspace{-10mm}
\caption{Upper panel: The ratio of the number of stars in the
field-star IMF relative to the stellar IMF (for $\alpha=2.35, 2.7,
3.2$, dotted, short-dashed and long-dashed lines, respectively) in the
mass interval $0.8\le m/M_\odot \le 8$ (the relative number of white
dwarfs) in dependence of the cluster mass-function power-law exponent
$\beta$.  Lower panel: The same except for $8\le m/M_\odot \le m_{\rm
max*}=150\,M_\odot$ (the relative supernova rate). The normalisation
of the IMFs is as in Fig.~\ref{fig:ex1}. Note that the panels have
different vertical scales.
\label{fig:rat}}
\end{figure}

Massey (2002) finds that massive stars that are not in OB associations
have $\alpha_{\rm iso}\approx4$. This steep IMF of isolated massive
stars has sometimes been taken to possibly imply a different mode of
star formation in isolated molecular cloud(lets) which may be arrived
at if the isolated interstellar medium has a different equation of
state (Spaans \& Silk 2000). An alternative may be dynamical ejections
of massive stars from cluster cores with high velocities (Clarke \&
Pringle 1992). Higher-mass stars typically have lower ejection
velocities and thus cannot spread as far into the field as less
massive stars.  The results obtained here alleviate this problem of
the isolated massive stars by allowing a steeper field-star IMF
(Fig.~\ref{fig:avsb}).

\acknowledgments

This research has been supported by DFG grants KR1635/3 and KR1635/4.



\clearpage


\end{document}